\documentclass[11pt]{article}
\usepackage{epsfig}
\usepackage{amssymb}
\begin{document}

\begin{center}
{\LARGE {\bf
The ANTARES Neutrino Project: Status Report
}}
\end{center}

\normalsize{
\begin{center}
Igor Sokalski~\footnote{INFN/Bari,
Via Amendola 173, I-70126 Bari, Italy, email: {\it Igor.Sokalski@ba.infn.it}} 
on behalf of the ANTARES Collaboration\footnote{The author list can be found 
at the end of this paper}\\
\end{center}
}

\begin{abstract}
The ANTARES project aims to build a deep underwater Cherenkov neutrino 
telescope in the Mediterranean Sea. Currently the experiment is in the 
construction phase and has recently achieved two important milestones. The 
electro-optical cable to shore and the junction box that will distribute power
to detector strings and allow data transmission have been deployed at the sea 
floor. A prototype string and a string for environmental parameter measurement
have been deployed, connected to the cable using a manned submarine. Data have
been sent to shore. The final ANTARES detector consisting in 12 strings each 
equipped with 75 photomultiplier tubes is planned to be fully deployed and 
taking data by the end of 2006.
\end{abstract}

\vspace{-84mm}
\begin{center}
{\it {\large Talk given at 4th International Conference on Non-Accelerator New 
Physics,\\ Dubna, Russia, June 23--28, 2003.
}}
\end{center}

\vspace{62mm}

\section{Introduction}
\label{sec:intro}

Neutrino is an attractive tool for astrophysical investigations since 
interacting weakly they can escape from the source and travel large distances 
to the Earth without interaction and without deflection by magnetic fields. 
Nevertheless, due to the same property, large volume neutrino detectors are 
needed. ANTARES is one of the several on-going projects [1-6] on 
underwater/ice neutrino telescopes. Given the presence of AMANDA at the South 
Pole, a detector in the Mediterranean will allow to cover the whole sky.
The ANTARES Collaboration ({\bf A}stronomy with a {\bf N}eutrino 
{\bf T}elescope and {\bf A}byss environmental {\bf RES}earch) was formed in 
1996 and currently joins about 200 scientists and engineers from France, 
Germany, Italy, Russia, Spain, The Netherlands  and the United Kingdom. The 
project aims to detect atmospheric and extraterrestrial neutrinos with 
energies above $E_{\nu} \sim$\,10\,GeV by means of the detection of the 
Cherenkov light that is generated in water by charged particles which are 
produced in $\nu N$ interactions. After extensive R\&D program the 
collaboration moved into construction of a detector in the 
Mediterranean Sea at 2400 m depth, 50 km off-shore of La Seyne sur Mer, near 
Toulon (42$^{\circ}$50$^{'}$\,N, 6$^{\circ}$10$^{'}$\,E).

\section{R\&D stage}
\label{sec:rd}

In 1996-99 an intense R\&D program was performed. The deployment and recovery 
technologies, electronics and mechanical structures were developed and tested
with more than 30 deployments of autonomous strings. The environmental 
properties at the detector site were investigated [7,8].

Concerning the optical backgrounds it was found that baseline 1\,p.e.-counting
rate of $\sim$60\,kHz is measured by a 10$^{''}$ PMT.
The counting rate increases during short bursts up to several MHz due to 
bioluminiscence. These bursts lead to a dead-time of less than 5\% per each 
PMT. However, the long-term measurements that were performed with the 
so-called 'prototype string' in 2003 (see below) showed that these rates and 
the burst fraction are sometimes essentially higher (Fig.\,1). The 
experimental work to understand the differences between previous results with 
autonomous mooring lines and the prototype string is in progress. Perhaps, to 
suppress the high background harder cuts will have to be applied which will 
slightly increase the energy threshold without altering the detection 
efficiency of $>$100\,GeV neutrino events. Light transmission loss for glass 
containers that house PMTs was found strong in long-term tests for up-looking 
surfaces. It led to the decision to turn all PMTs downward. Signal loss due to
bio-fouling and sedimentation was measured to be 1.6\% after 8 months at 
equator of glass sphere saturating with time. The optical properties of water 
at the experiment site were measured during several years. The effective 
attenuation length varies in a range 48\,m\,\,$<L_{att}<$\,\,61\,m while 
scattering length is  $L_{scatt}>$\,\,200\,m for blue light 
($\lambda=$\,466 nm). Only 5\% of the photons emitted by an isotropic source 
located 24\,m from PMT are collected out of a 10\,ns time window being delayed
due to scattering. This allows a good time resolution needed for event 
reconstruction. 

ANTARES R\&D program culminated with deployment and 8 month operation of a 
350\,m length 'demonstrator string' (November 1999 - July 2000) instrumented 
with 7 PMTs at a depth of 1100\,m, 40 km off the coast of Marseille. The string
was controlled and read out via 37\,km-long electro-optical cable connected to
the shore station. It allowed to test the deployment procedure with a 
full-scale string, positioning 
system and collect $\sim$5$\cdot$10$^{4}$ seven-fold coincidences from 
atmospheric muons. Relative distances were measured with an accuracy of 
$\sim$5\,cm and accuracy of absolute positioning was $\sim$1\,m. The angular 
distribution of atmospheric muons was reproduced and the fraction of 
multi-muon events was found to be $\sim$50\% which is in agreement with 
expectation for such a shallow depth as 1100\,m.  

\section{ANTARES detector}
\label{sec:0.1}

After this R\&D experience, the collaboration moved to the next stage:
construction of a 12-string detector [2] which can be considered 
as a step toward a 1\,km$^{3}$ detector (Fig.\,2). Strings are anchored at the
sea floor and held taut by buoys. Each string is instrumented with 75 optical 
modules (OMs) [9] containing 10$^{``}$ Hamamatsu R7081-20 PMTs housed in glass
spheres. OMs are grouped in triplets at 25 levels separated by 14.5\,m. 3 PMTs
in each triplet are oriented at 45$^{\circ}$ to the nadir. Strings are 
separated from each other by $\sim$70\,m. All the strings are connected to a 
Junction Box (JB) by means of electro-optical link cables. The JB is connected
to the shore station by a 50\,km length 48-fiber electro-optical cable. 
Undersea connections are performed with a manned submarine. PMT signals are 
processed by Analogue Ring Samplers ASIC which measure the arrival time and 
charge for 1\,p.e.-pulses (99\% of the pulses) and perform wave form 
digitization for larger amplitudes. Digitized data from each OM are sent to 
shore ($\sim$1 GB/s/detector). The data flow is reduced down to $\sim$1 MB/s 
by means of an on-shore data filter [10]. 100 PC farm is foreseen on shore to 
process and collect the data. The telescope will be complemented with an 
instrumentation string for hydrological parameter measurements and for 
calibration purposes. The deployment of the detector is planned for 2004-2006.

The important milestones that have been achieved by the collaboration are: 

\begin{itemize}
\item[{\it 1)}] the electro-optical cable connecting detector and shore station was 
deployed in October 2001; 
\item[{\it 2)}] the industrial production of 900 OMs started
in April 2002; 
\item[{\it 3)}] since December 2002 the JB is in communication with the
shore station; 
\item[{\it 4)}]  in December 2002 and February 2003 the 'prototype 
instrumentation string' and the 'prototype detection string' (equipped with 15
OMs) were successfully deployed [11] (recovered in May and July, 2003, 
respectively); 
\item[{\it 5)}] in March 2003 both strings were connected to JB with 
the Nautile manned submarine and data taking started. 
\end{itemize}

The aim of the 
deployment and operation of two prototype strings were to test all the 
components of the future detectors in their final design. 
Mechanical problems occurred: 1 fiber
for clock signal transmission was
found broken and 1 connector leaked. After strings recovery it was found that
these problems occurred due to manufacturers who changed design without 
notification. Solutions have been found for the final detector design and 
severe quality control will be applied.

The detailed description of ANTARES physics performance can be found in [12].
The angular resolution of the 12-string detector (Fig.\,3) is about 
0.2$^{\circ}$ for $E_{\nu} \ge$\,100\,TeV where it is limited only by PMT TTS 
and light scattering and $\sim$0.5$^{\circ}$--1$^{\circ}$ at 
$E_{\nu} \sim$\,0.1--10\,TeV where accuracy is dominated by $\nu - \mu$ 
kinematics. Energy resolution (Fig.\,4) improves at high energies: dispersion 
of the $\log_{10}(E_{rec}/E_{t})$ distribution (where $E_{t}$ is the true 
energy and $E_{rec}$ is the reconstructed energy, respectively) is around  
$\sigma \approx$\,0.5 at $E_{\nu} \sim$\,5\,TeV and $\sigma \approx$\,0.3 for 
$E_{\nu} \ge$\,100\,TeV. Effective area for muons grows from 
$A_{eff}=$\,0.01\,km$^{2}$ at $E_{\nu} =$\,1\,TeV to  
$A_{eff}=$\,0.06\,km$^{2}$ at $E_{\nu} =$\,10\,PeV. The sensitivity of the 
detector to diffuse neutrino fluxes achieved by rejecting the background with 
an energy cut of $E_{cut}=$\,50\,GeV  allows to  reach Waxmann \& Bahcall 
limit [13] in 3 years. The ANTARES sensitivity for point-like source searches 
(90\% C.L.) assuming $E^{-2}$ differential $\nu$ flux is in the range 
4$\div$50$\cdot$10$^{-16}$\,cm$^{-2}$\,s$^{-1}$ after 1 yr which gives a real 
hope to detect a signal from the most promising sources (e.g., galactic 
microquasars [14]). The ANTARES potential for WIMP searches is high enough to 
improve existing experimental upper limits on $\nu$-induced muon fluxes from 
neutralino annihilation in the Sun and on relativistic magnetic monopole flux 
obtained by other detectors by an order of magnitude.

\section{Conclusions}
\label{sec:conc}

The construction of the ANTARES detector is underway. It is
planned to be fully deployed and start to take data by the end of 2006. 
Calculations based on the data on environmental conditions at the experiment 
site and on studied properties of electronic components shows that predicted 
sensitivity of the detector to diffuse neutrino fluxes, point-like neutrino 
searches and WIMP searches is better by several orders of magnitude compared 
to data published by other experimental groups. The deployment of the 
ANTARES neutrino telescope can be considered as a step toward 
the deployment of a 1 km$^3$ detector in the Mediterranean Sea.

\begin{center}
The ANTARES Collaboration:
\end{center}

\vspace{-2.0mm}

\begin{center}
{\footnotesize {\sf
J.~A.~Aguilar$^{1}$,             
H.~Alleyne$^{2}$,                
F.~Ameli$^{3}$,                  
P.~Amram$^{4}$,                  
M.~Anghinolfi$^{5}$,             
G.~Anton$^{6}$,                  
S.~Anvar$^{7}$,                  
F.~E.~Ardellier-Desages$^{7}$,   
E.~Aslanides$^{8}$,              
J.-J.~Aubert$^{8}$
% ...
\\[1mm]       

\strut
}}
\end{center}

\vspace{-7.5mm}

{\tiny
{\sf

\noindent
{\bf 1} - IFIC, Edificios Investigaci\'on 
    de Paterna, CSIC, Universitat de Val\`encia, Apdo. de Correos 22085, 
    46071 Val\'encia, Spain; 
{\bf 2} - University of Sheffield, Department of Physics and Astronomy, 
    Hicks Building, Hounsfield Road, Sheffield S3 7RH, United Kingdom; 
{\bf 3} - Dipartimento di Fisica dell'Universit\`a "La Sapienza" e Sezione INFN, 
    P.le Aldo Moro 2, 00185 Roma, Italy; 
{\bf 4} - LAM, CNRS/INSU,  
    Universit\'e de Provence Aix-Marseille I, Traverse du Siphon -- 
    Les Trois Lucs, BP 8, 13012 Marseille Cedex 12, France; 
{\bf 5} - Dipartimento di Fisica dell'Universit\`a e Sezione INFN, Via Dodecaneso 
    33, 16146 Genova, Italy; 
{\bf 6} - Friedrich-Alexander Universit\"at 
    Erlangen-N\"urnberg, Physikalisches Institut, Erwin-Rommel-Str. 1, 91058 
    Erlangen, Germany;  
{\bf 7} - DSM/DAPNIA, CEA/Saclay, 91191 Gif Sur Yvette Cedex, France; 
{\bf 8} - CPPM, CNRS/IN2P3 
    Universit\'e de la M\'editerran\'ee Aix-Marseille II, 163 Avenue de 
    Luminy, Case 907, 13288 Marseille Cedex 9, France; 
{\bf 9} - Dipartimento di Fisica dell'Universit\`a e Sezione INFN, Viale Berti 
    Pichat 6/2, 40127 Bologna, Italy; 
{\bf 10} - Dipartimento Interateneo di Fisica e Sezione INFN, Via E. Orabona 4, 
     70126 Bari, Italy; 
{\bf 11} - GRPHE, 
     Universit\'e de Haute Alsace, 61 Rue Albert Camus, 68093 Mulhouse Cedex, 
     France; 
{\bf 12} - COM, CNRS/INSU Universit\'e de la 
     M\'editerran\'ee Aix-Marseille II, Rue 
     de la Batterie des Lions, 13007 Marseille, France; 
{\bf 13} - NIKHEF, Kruislaan 409, 1009 SJ Amsterdam, The Netherlands;  
{\bf 14} - University of Leeds, Department of Physics and Astronomy, Leeds LS2 9JT, 
     United Kingdom; 
{\bf 15} - Dipartimento di Fisica ed Astronomia dell'Universit\`a e Sezione INFN, 
     57 Corso Italia, 95129 Catania, Italy; 
{\bf 16} - UMR G\'eoScience Azur, Observatoire Oc\'anologique de Villefranche, BP48, 
     Port de la Darse, 06235 Villefranche-sur-Mer Cedex, France; 
{\bf 17} - IFREMER (Brest), BP 70, 29280 Plouzan\'e, France; 
{\bf 18} - ITEP, B.~Cheremushkinskaya 25, 117259 Moscow, Russia; 
{\bf 19} - IFREMER (Toulon/La Seyne Sur Mer), Port Br\'egaillon, Chemin 
     Jean-Marie Fritz, 83500 La Seyne Sur Mer, France; 
{\bf 20} - IReS (CNRS/IN2P3), Universit\'e 
      Louis Pasteur, BP 28, 67037 Strasbourg Cedex 2, France; 
{\bf 21} -  Universit\'e Paris VII, Laboratoire APC, 4 place Jussieu, Tour 33, 
      Rez de chausse 75252 Paris Cedex 05, France; 
{\bf 22} - INFN -- LNS, Via S. Sofia 44, 
     95123 Catania, Italy; 
{\bf 23} - ISS, 76900 Bucharest, Romania.
}

\strut
}

\vspace{-0mm}

\newpage

\begin{center}
FIGURES
\end{center}

\vspace{18mm}

\begin{figure}[h]
\begin{center}
\includegraphics[width=14.8cm]{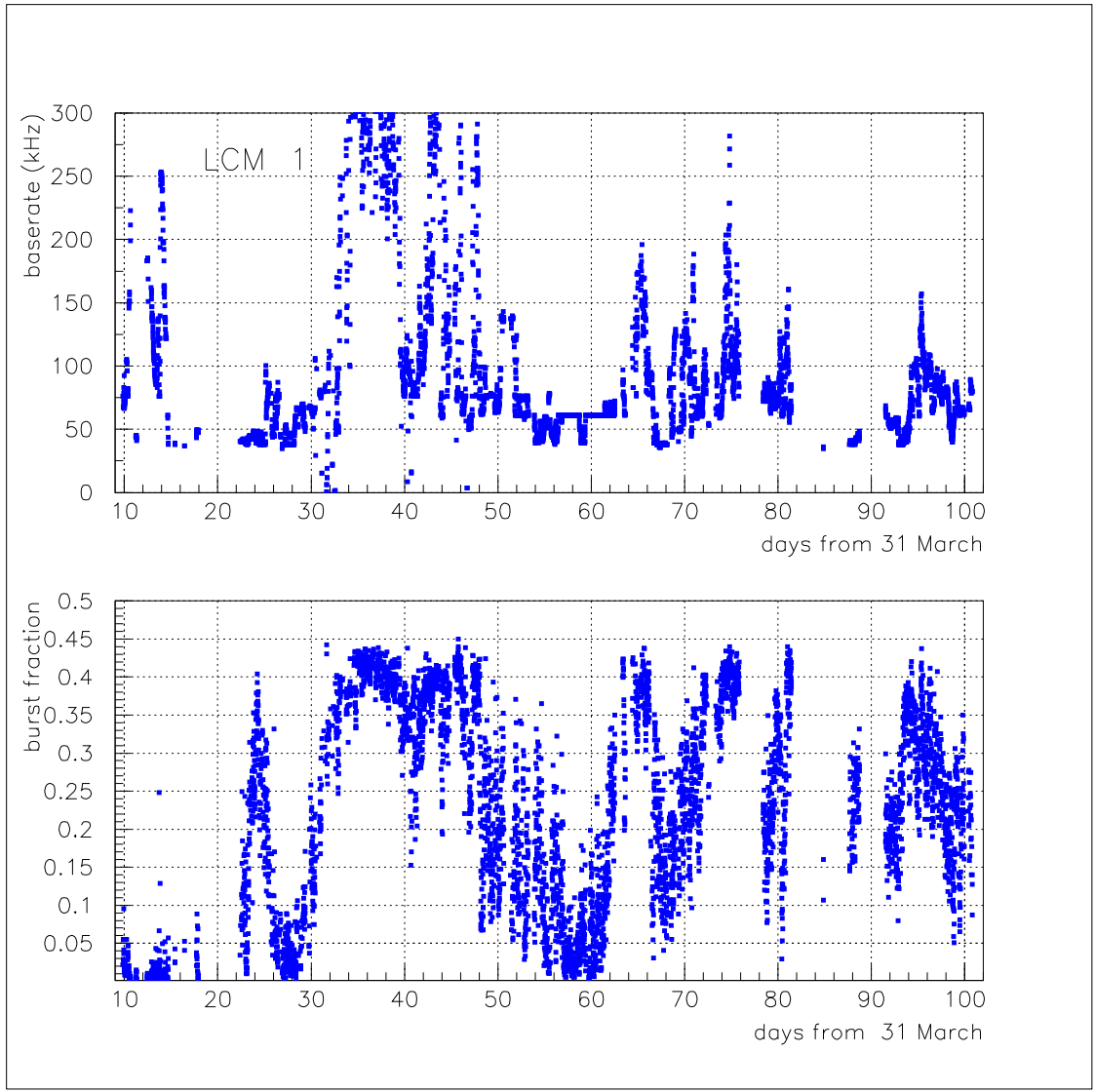}
\caption{
Summary of counting rate in 3 PMTs during 65 days of 
'prototype string' operation in April--May, 2003. 
Top figure: the average baseline rate. Bottom
figure: the fraction of time the rate is significantly higher than this average
baseline rate (burst fraction).
}
\end{center}
\label{fig1}
\end{figure}

\begin{figure}[h]
\begin{center}
\includegraphics[width=15.8cm]{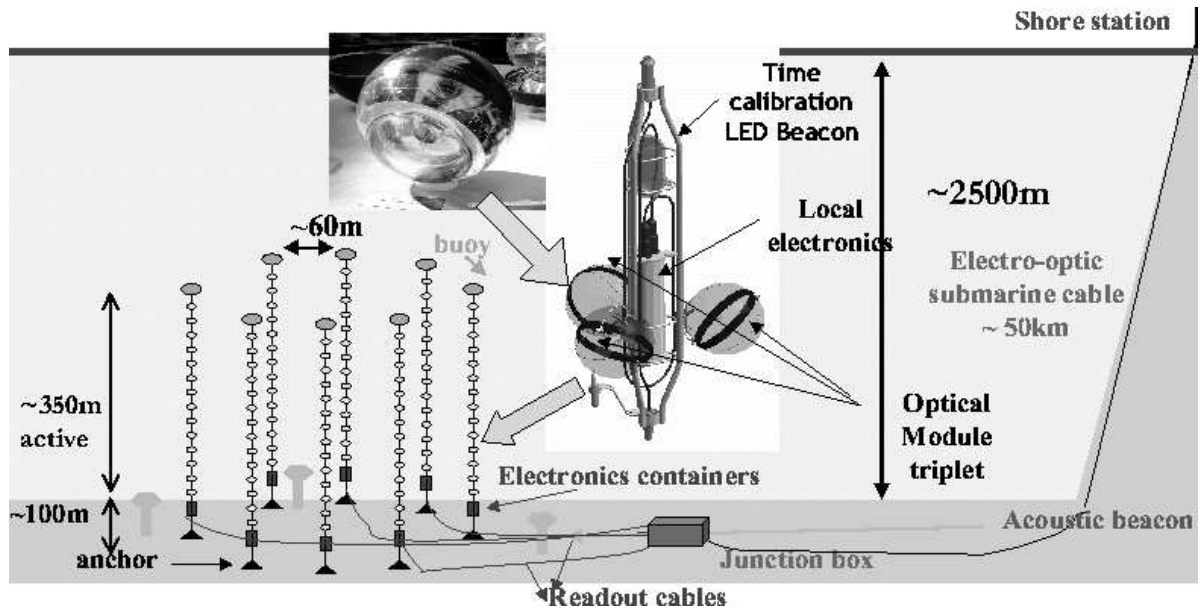}
\caption{
Schematic view of the ANTARES 12-string detector.
}
\end{center}
\label{fig2}
\end{figure}

\begin{figure}[h]
\begin{center}
\includegraphics[width=14.0cm]{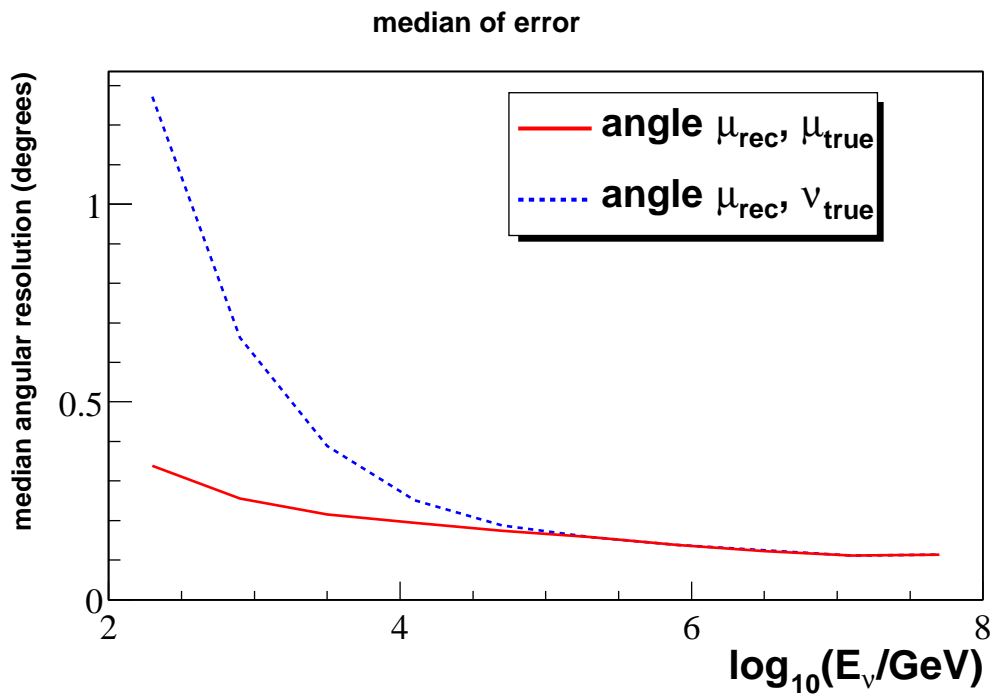}
\caption{
Angular resolution of the ANTARES detector versus $E_{\nu}$: median
of the distribution of the angle in space between the reconstructed muon
track and true muon track (solid) or the parent neutrino track (dashed). 
}
\end{center}
\label{fig3}
\end{figure}

\begin{figure}[h]
\begin{center}
\includegraphics[width=15.6cm]{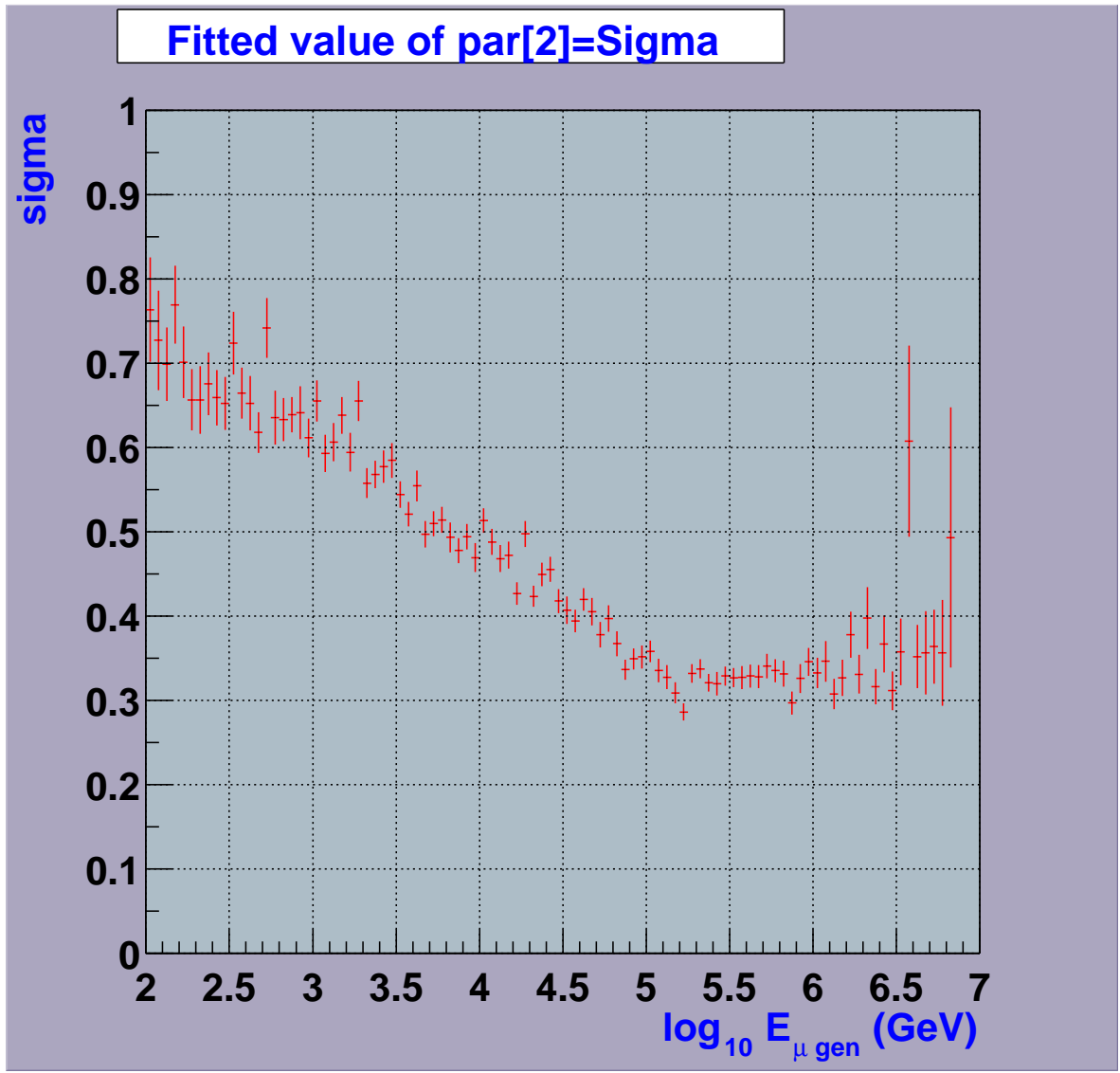}
\caption{
Energy resolution of the ANTARES detector: sigma of the
distributions of $\log_{10} (E_{rec} / E_{gen})$ (where $E_{rec}$ is
reconstructed muon energy and $E_{gen}$ is generated muon energy)
versus generated energy.
}
\end{center}
\label{fig4}
\end{figure}

\end{document}